\documentclass[a4paper]{jpconf}
\usepackage{graphicx}
\usepackage{tikz}
\usetikzlibrary{shapes,arrows,chains}
\begin{document}
\title{MEtop - a top FCNC event generator}

\author{Rita Coimbra$^1$, Ant\'onio Onofre$^2$, Rui Santos$^{3,4}$ and Miguel Won$^{1,4}$}

\address{$^1$ LIP / Departamento de F\'{\i}sica, Universidade de Coimbra, 3004-516 Coimbra, Portugal}
\address{$^2$ Departamento de F\'{\i}sica, Universidade do Minho, 4710-057 Braga, Portugal}
\address{$^3$  Instituto Superior de Engenharia de Lisboa - ISEL, Rua Conselheiro Em\'\i dio Navarro 1, 1959-007 Lisboa, Portugal }
\address{$^4$ Centro de F\'\i sica Te\' orica e Computacional, Faculdade de Ci\^encias, Universidade de Lisboa, Av. Prof. Gama Pinto 2, 1649-003 Lisboa, Portugal}

\ead{rita.coimbra@coimbra.lip.pt, Antonio.Onofre@cern.ch, rsantos@cii.fc.ul.pt, miguel.won@coimbra.lip.pt}

\begin{abstract}
In this work we present a new Monte Carlo generator for Direct top and Single top production via flavour-changing neutral currents (FCNC). 
This new tool calculates the cross section and generates events with Next-to-Leading order precision for the Direct top process
and  Leading-Order precision for all other FCNC single top processes. A set of independent dimension six FCNC operators has been implemented
- including four-fermion operators - where at least one top-quark is present in the interaction. 
\end{abstract}

\section{Introduction}
With the new experimental data from the Large Hadron Collider (LHC) one will be able to scrutinise the Standard Model (SM) 
boundaries with unprecedented  precision.
 Flavour-changing neutral currents (FCNC) top decays such as $t \to V \, q$, where $V=Z, \gamma,g$ and $q=u,c$ are
 highly suppressed  in the SM due to the well known Glashow-Iliopoulos-Maiani (GIM) mechanism~\cite{GIM}. top-quark FCNC 
 branching ratios are of the order of $10^{-14}$ and $10^{-12}$ for the electroweak and strong case respectively. 
 However, some versions of two-Higgs doublet models (2HDMs) predict quite larger values. In fact, for some regions of the 2HDM parameter
 space the BR($t \to g \, q$) could reach values of the order of $10^{-4}$~\cite{AguilarSaavedra:2004wm}, which is eight orders of magnitude 
 above the SM value. 
%
The high statistics expected in LHC data forces an increase in the precision of all relevant theoretical calculations. Hence, 
at least Next-to-Leading (NLO) calculations should be made available whenever possible. The experimental searches 
performed both at LHC and at the Tevatron on the direct top FCNC process have already used NLO cross sections from~\cite{Liu:2005dp}. 
However, the NLO calculation was used only as a normalization factor and  the events were generated using LO generators like 
TopRex~\cite{Slabospitsky:2002ag} or Protos~\cite{PROTOS}. 
In this work we present a new Monte Carlo tool that generates FCNC  direct top events with NLO precision. The FCNC interactions 
were implemented via an effective lagrangian~\cite{buch} where a complete set of dimension six operators was used. 
The events are generated in the Les-Houches Event (LHE) format~\cite{Alwall:2006yp} and  can therefore 
be easily interfaced with shower algorithms such as the one in Pythia~\cite{PYTHIA6}. Since the generator calculates the total inclusive 
 cross section, the FCNC single top process was also included and can be generated independently. Not only strong FCNC interactions,   
but also electroweak and four-fermions dimensions six operators were included. Each operator can be turned on independently allowing 
for independent studies of the different Lorentz structures.  In  section~\ref{sec:effective}  the effective formalism
is presented while in section~\ref{sec:processes} we show the physical processes already included in MEtop. 
In section~\ref{sec:NLO} we give a description of the algorithm used by the Monte Carlo generator in order to obtain a final NLO result. 
In section~\ref{results} we discuss a few selected results and in section~\ref{sec:conclusion} we sum up our conclusions.

\section{Effective operators}
\label{sec:effective}
In the effective lagrangian formalism we assume that the SM is no more than a low energy limit of some more general theory to be revealed at TeV scale. 
In this context, the top-quark FCNC interaction is seen at low energies as a contact interaction that can be parametrized with higher dimension effective operators. 
The complete lagrangian is written as
\begin{equation}
{\cal L} = {\cal L}^{SM} +\frac{1}{\Lambda} {\cal O}^5 + \frac{1}{\Lambda^2} {\cal O}^6 + {\cal O}( \frac{1}{\Lambda})
\label{eqn:lgrgeral}
\end{equation}
where we have truncated the expansion series at dimension $d=6$.
We follow the formalism in~\cite{buch}, where the SM symmetries are imposed and all operators are built using the already known SM fields. 
This results in a reduced list of the possible dimension six operators. Additionally, baryon and lepton number conservation is imposed, which restricts
the operators to dimension six only. Even imposing the SM symmetries, the number of all possible FCNC operators is quite vast. However, the list can
be reduced by using appropriate equations of motion and Fierz transformations applied to the physical operators we are dealing with. The final minimal 
complete set of FCNC operators for top-quark physics was obtained in~\cite{Ferreira:2005dr,Ferreira:2006xe, AguilarSaavedra:2008zc, Grzadkowski:2010es}. 
With this procedure, a single operator survives in the strong sector~\cite{AguilarSaavedra:2008zc}. Following the notation in~\cite{buch} this operator can be written as
\begin{equation}
{\cal O}^{ij}_{uG \phi} \, = \,  \bar{q}^{i}_L \, \lambda^{a} \, \sigma^{\mu\nu} \, u^{j}_R \, \tilde{\phi} \, G^{a \mu \nu}  \, \label{eq:op1}
\end{equation}
where $G^a_{\mu\nu}$ is the gluonic field tensor, $u^i$ right-handed $u$ or $c$-quark field and $q^i$ is the left-handed quark doublet. 
The same operator also appears in the literature as a dimension 5 operator
\begin{equation}
i \kappa_i \, \frac{g_s}{\Lambda} \bar{q_i} \lambda^a \sigma^{\mu\nu} (f_i+h_i \, \gamma_5) t  G_{\mu\nu}^a \quad
\label{eq:op1},
\end{equation}
where $ \kappa_i$ is a real parameter, $i=u,c$, $g_s$ is the strong coupling constant and $f_i$ and $h_i$ are complex numbers which respect $|f_i|^2+|h_i|^2=1$.
This operator is responsible for the effective FCNC vertices $tgu$ and $tgc$. The minimum set of operators in the electroweak sector is
\begin{equation}
{\cal O}^{ij}_{uB \phi}  \, = \, \bar{q}^i_L \, \sigma^{\mu\nu} \, u^j_R \, \tilde{\phi} \,B_{\mu\nu} \, , \quad
{\cal O}^{ij}_{uW \phi} \, = \, \bar{q}^i_L \, \tau_{I} \, \sigma^{\mu\nu} \, u^j_R \, \tilde{\phi} \, W^{I}_{\mu\nu} \,  ,
\label{eq:op3}
\end{equation}
\\[-0.7cm]
\begin{equation}
{\cal O}^{ij}_{\phi u}  \, = \, i  \, (\phi^{\dagger}  D_{\mu} \phi) \, (\bar{u}^i_R \,  \gamma^{\mu}  \,  u^j_R) \, , 
\\
\label{eq:op4}
\end{equation}
\\[-0.7cm]
\begin{equation}
{\cal O}^{(1),ij}_{\phi q} \, = \, i \,  (\phi^{\dagger} D_{\mu} \phi) \, (\bar{q}^i_L   \, \gamma^{\mu}  \,  q^j_L)  \, , \quad
{\cal O}^{(3),ij}_{\phi q} \, = \, i \, (\phi^{\dagger} \, \tau_{I}  \, D_{\mu} \phi)  \,   (\bar{q}^i_L \, \gamma^{\mu} \, \tau_{I}  \,  q^j_L)  \, ,
\label{eq:op5}
\end{equation}
\\[-0.7cm]
\begin{equation}
{\cal O}^{ij}_{u \phi} \, =   (\phi^{\dagger}  \phi) \, (\bar{q}^i_L \,  u^j_R \, \tilde{\phi})  \, , \quad
\label{eq:op6}
\end{equation}
where $B^{\mu \nu}$ and $W^{I}_{\mu\nu}$ are the $U(1)_Y$ and $SU(2)_L$ field 
tensors, respectively. The field $\phi$ is the usual Higgs boson doublet. In addition to this electroweak set of operators,
 we have included subsets of four-fermion operators which are described in ~\cite{metop}.

\section{Physical Processes}
\label{sec:processes}
\begin{figure}[h!]
\begin{center}
\includegraphics[scale=0.7]{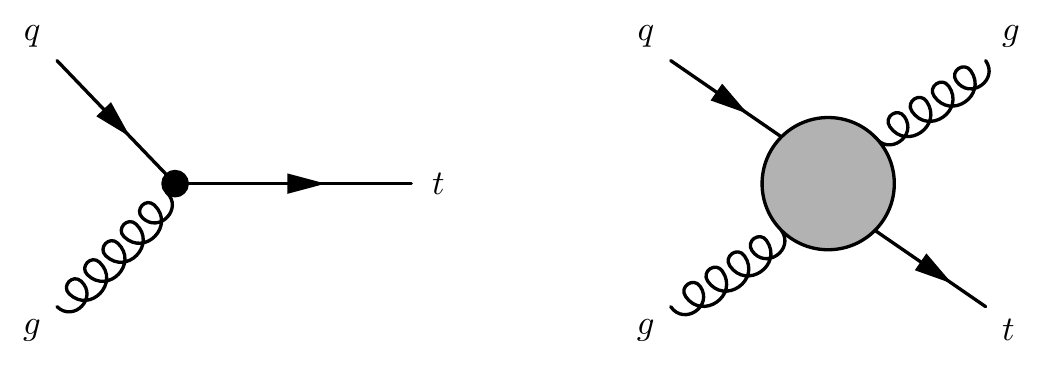}
\caption{FCNC leading order direct top production and top + gluon production at the parton level. 
Only FCNC strong operators contribute to the process.}
\label{fig:Dtop}
\end{center}
\end{figure}

\begin{figure}[h!]
\begin{center}
\includegraphics[scale=0.7]{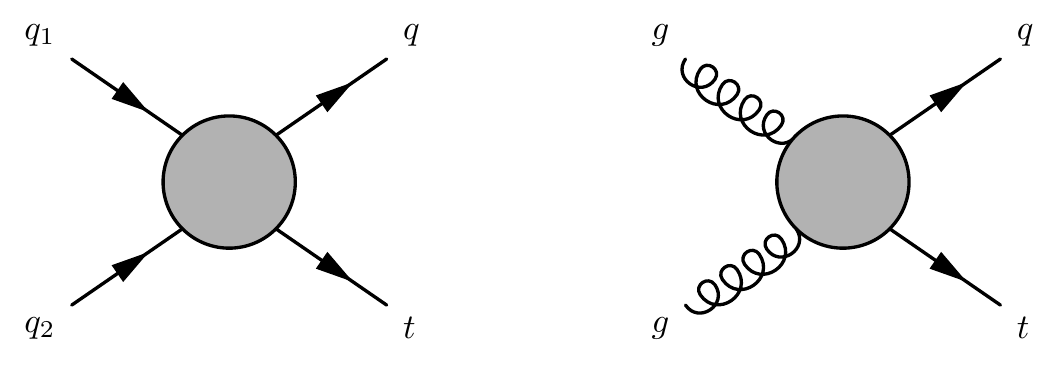}
\caption{FCNC leading order top + quark production at the parton level. FCNC strong and 
electroweak operators contribute to the process together with 4F operators.}
\label{fig:Lqtop}
\end{center}
\end{figure}

The physical processes included in MEtop are FCNC direct top production and FCNC single top production. Direct top is available both at LO and at NLO while 
FCNC single top is availble only at LO. All processes can be generated for both  $pp$ and $p\bar{p}$ hadron colliders. Furthermore, the full top decay chain 
is included so that spin correlations are preserved. Also, in order to  perform spin correlations studies, a on/off switch for the top decay chain was made available. 
In figures~\ref{fig:Dtop} and~\ref{fig:Lqtop} we show the generic Feynman diagrams for the physical processes implemented in MEtop. The production process depicted
in figure~\ref{fig:Dtop} has only strong FCNC operator contributions. Born level FCNC direct top is shown in figure~\ref{fig:Dtop} (left) together with the $t+g$ process (right). 
In figure~\ref{fig:Lqtop} we present the diagrams for the LO $t+q$ production process from gluon fusion (right) and with quarks in the initial state (left) . The former 
is only affected by strong FCNC operators while the later has contributions from the strong, electroweak and four-fermion sectors.
\section{NLO approximation}
\label{sec:NLO}
\begin{figure}[h!]
\begin{center}
\includegraphics[scale=0.7]{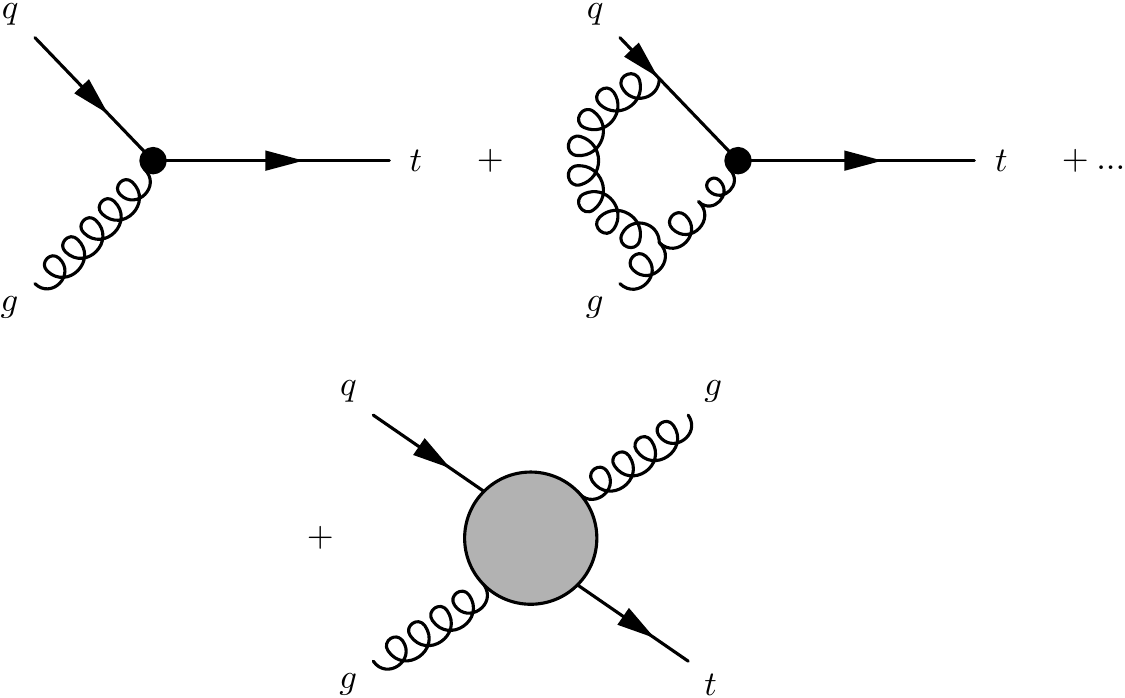}
\caption{Inclusive FCNC direct top production at NLO in QCD.}
\label{fig:DtopNLO}
\end{center}
\end{figure}

As mentioned in the previous section, MEtop is able to generate FCNC direct top events with NLO precision. The implementation 
was performed by  adopting an NLO effective approximation~\cite{Boos:2006af}. 
A general NLO calculation of an n-particles final state process can be represented by
\begin{equation}
d\sigma_{NLO} = d\sigma_B d\Phi_n + d\sigma_{V} d\Phi_n + d\sigma_{R} d\Phi_{n+1},
\label{eqn:sigmaNLO}
\end{equation} 
where $B$ stands for Born, $V$ for Virtual and $R$ for Real. It is well known that the virtual term has infrared divergences 
that will only be cured by the inclusion of the real radiation term $\sigma_R$. We show in figure~\ref{fig:DtopNLO} the 
class of diagrams contributing to NLO direct top production. The first process $qg \to t$ correspond to the Born term.  
The second, $qg \to t$, corresponds to the Virtual term while the third, $qg \to gt$, is the real radiation term. This last term
gives rise to an infrared divergence which correspond to a low $P_T$ and/or to a collinear external gluon.  
Kinematically, the divergence can be parametrized by the gluon transverse momentum, $P_T$, and it obviously occurs in the limit $P_T \to 0$.
 In order to avoid the divergence one could just impose a $P_T$ cut.  The resulting region has to be excluded from the event space
 generation with some care - since the problem arises only in the collinear/soft limit, we can fill this phase 
 space region with a branching mechanism. By doing so, we assume a collinear factorization, where the QCD radiation
  is emitted by one of the external legs of the Born configuration. This approximation can be translated by
\begin{equation}
|M_{t+g}|^2 d\Phi_{2}  \to  |M_{t}|^2 d\Phi_{1} \frac{\alpha_S}{2\pi} \frac{dt}{t} P_{q,qg} (z) dz \frac{d\phi}{2\pi}
\label{eqn:Shower}
\end{equation}
where $M_{t+g}$ and $M_{t}$ are the real radiation for ($t+g$) and  direct top amplitudes, respectively; $d\Phi_{i}$ is the phase space
for the $i$th body processes and $P_{q,qg} (z) $ is the 
Altarelli-Parisi splitting function. The variable $t$ is a resolution parameter which in our case is
the top quark transverse momentum $P_T^{top}$. The hard $P_T$ region will then be filled by using the appropriate transition amplitude and  
the low $P_T$ region by the use of a shower mechanism. In this NLO effective approximation~\cite{Boos:2006af}, the $P_T$ phase-space is 
then divided in two regions parametrized by $P_T^{top}$, which we call $P_T^{match}$.  This parameter plays the role of a matching parameter
where the criterion for a good transition is the smoothness of the final $P_T$ distribution.
The final cross section free of divergences  can be written as
\begin{equation}
\sigma_{NLO} = K \sigma_{B} (P_T^{PS} < P_T^{match}) + \sigma_{R}(P_T > P_T^{match}),
\label{eqn:kfactor}
\end{equation}
where $P_T^{PS}$ is the transverse momentum of the parton shower emission. 
The contribution from the virtual term $d\sigma_V$ in equation~\ref{eqn:sigmaNLO} is 
included via a K-factor applied to the born term and because born and virtual terms share the same $2 \to 1$ topology 
we assume they have similar kinematics.
It is from equation~\ref{eqn:kfactor} that events in MEtop are generated. Since MEtop is a parton level generator, the events must then be submitted
 to a parton shower simulator in order to include initial (ISR) and final (FSR) state radiation. Therefore, since these parton showers are based on
 a branching mechanism, we leave the collinear and/or soft $P_T$ phase space region to be filled by the corresponding parton shower. In order to avoid
  double counting we just have to ensure that the parton shower will not populate the $P_T > P_T^{match}$ region in the first emission. To prevent it,
   the shower mechanism must be performed with the PT-ordered scheme with the first emission starting at $P_T = P_T^{match}$. For the NLO direct top 
   process, MEtop assumes that the Pythia shower will be used with its PT-ordered scheme~\cite{Sjostrand:2004ef} flag on. Finally, we note that because 
   this is a $2 \to 1$ process, the shower mechanism implemented in Pythia will not add FSR but only ISR. This does not pose any problems because 
   FSR is negligible when compared to ISR due the large top-quark mass.
\section{The generation process and some results}
\label{results}
In figure 4 we show the diagram  flow of the MEtop generation process. It was drawn specifically for the NLO direct top case. 
The remaining processes follow the same flow except for the K-factor calculation step. The amplitudes used to evaluate 
the cross sections were generated with CalcHEP~\cite{Pukhov:1999gg}, whit Feynman rules derived with
 LanHEP \cite{Semenov:1998eb} and all integrations were performed using the Cuba Library~\cite{Hahn:2004fe}.
 The K-factors are calculated "on the fly" and therefore each sub process has its own normalization factor.
\tikzstyle{block} = [rectangle, draw, fill=blue!20, 
    text width=20em, text centered, rounded corners, minimum height=1.3em]
\tikzstyle{line} = [draw, -latex']
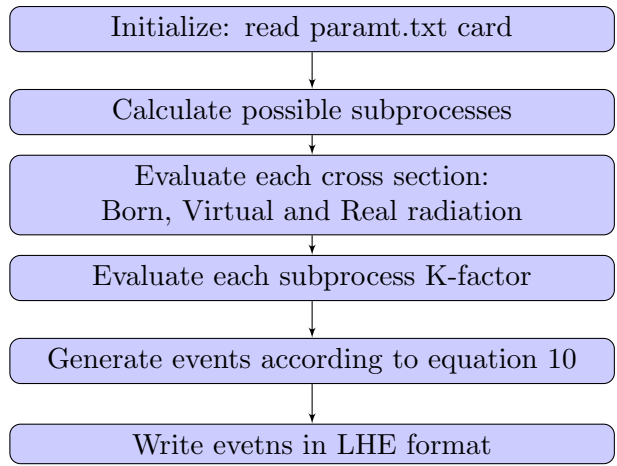
\begin{figure}[h!]
\begin{center}
\begin{tikzpicture}[node distance = 1.1cm, auto]
    \node [block] (init) {Initialize: read paramt.txt card};
    \node [block, below of=init] (identify) {Calculate possible subprocesses};
    \node [block, below of=identify] (evaluate) {Evaluate each cross section: Born, Virtual and Real radiation};
      \node [block, below of=evaluate] (kfactor) {Evaluate each subprocess K-factor};
      \node [block, below of=kfactor] (generate) {Generate events according to equation \ref{eqn:kfactor}};
      \node [block, below of=generate] (save) {Write evetns in LHE format};
    \path [line] (init) -- (identify);
    \path [line] (identify) -- (evaluate);
    \path [line] (evaluate) -- (kfactor);
     \path [line] (kfactor) -- (generate);
     \path [line] (generate) -- (save);

\end{tikzpicture}
     \label{flow:NLODtop}
\caption{MEtop flow cart for NLO Direct top generation.}
\end{center}
\end{figure}

\begin{figure}[h]
\begin{center}
\includegraphics[scale = 0.6]{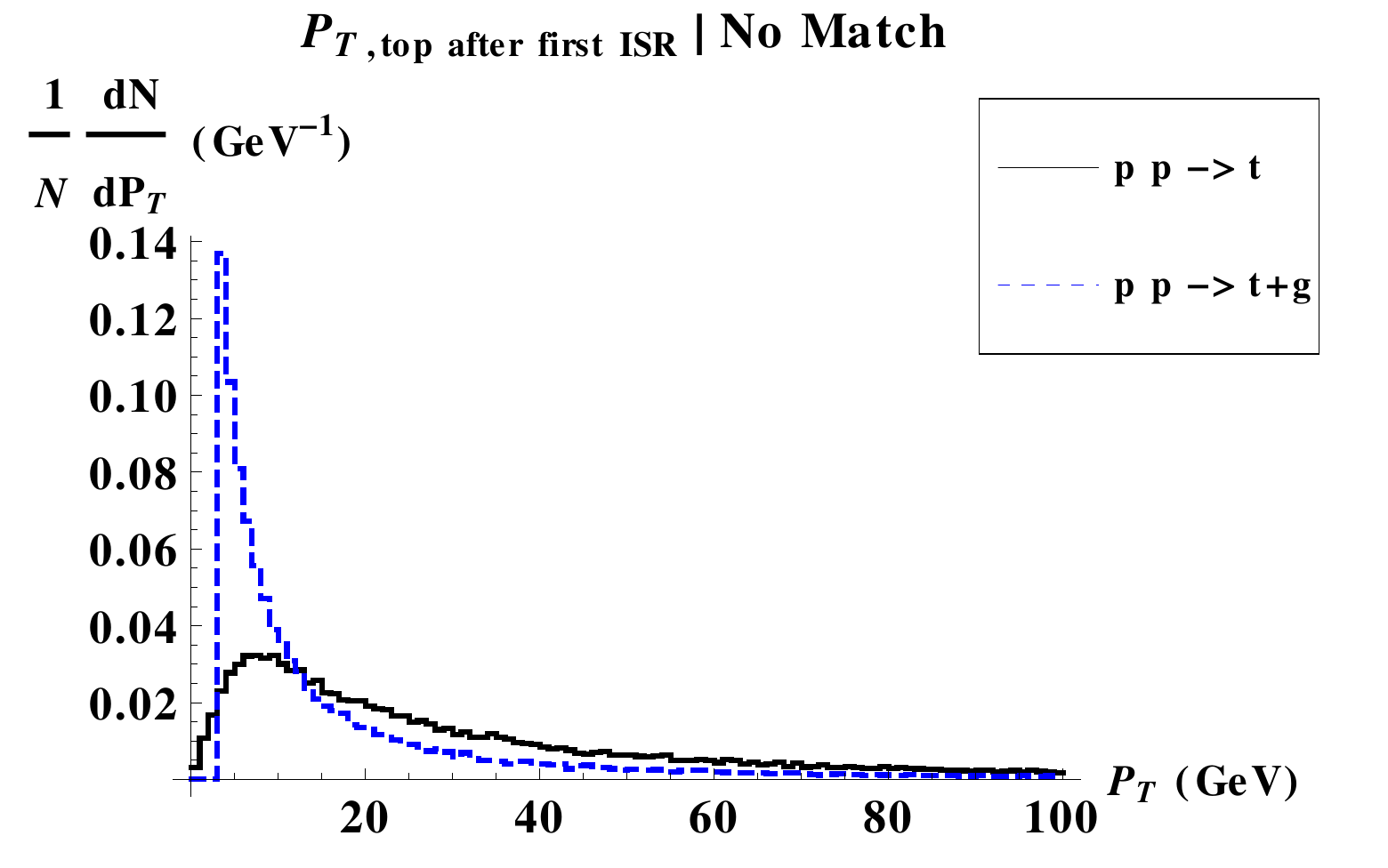}
\caption{$P_T$ distribution of the top quark for $\sqrt{s}$ = 7 TeV. The black solid line is for direct top production after the first branching in ISR,
 with starting scale of $m_{t}$. The blue dashed line is for the hard process top+gluon production.}
\label{fig:NoMatch}
\end{center}
\end{figure}

\begin{figure}[h!]
\centering
\hspace{-1.cm}\includegraphics[width=3.0in,angle=0]{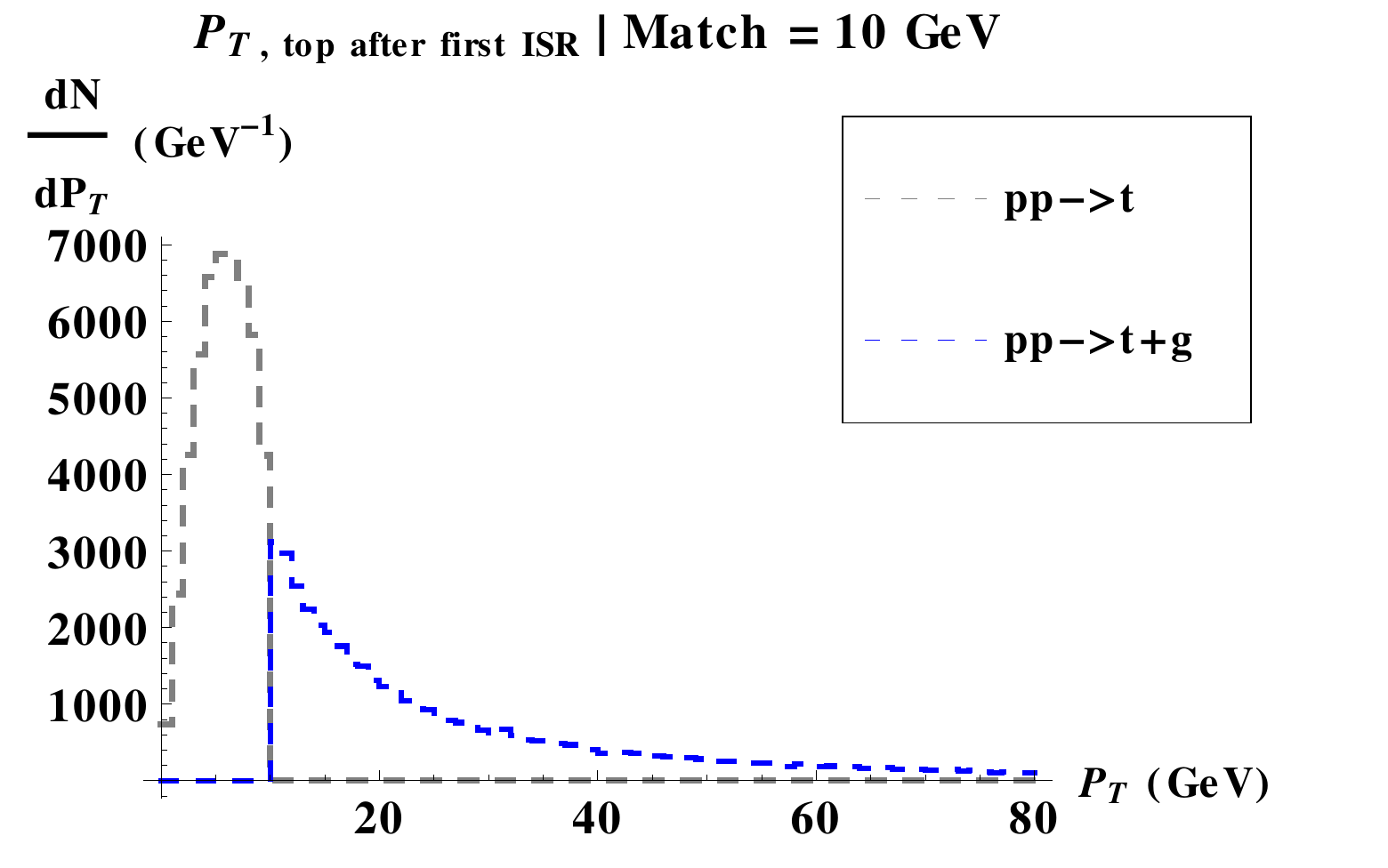}
\hspace{-.3cm}
\includegraphics[width=3.0in,angle=0]{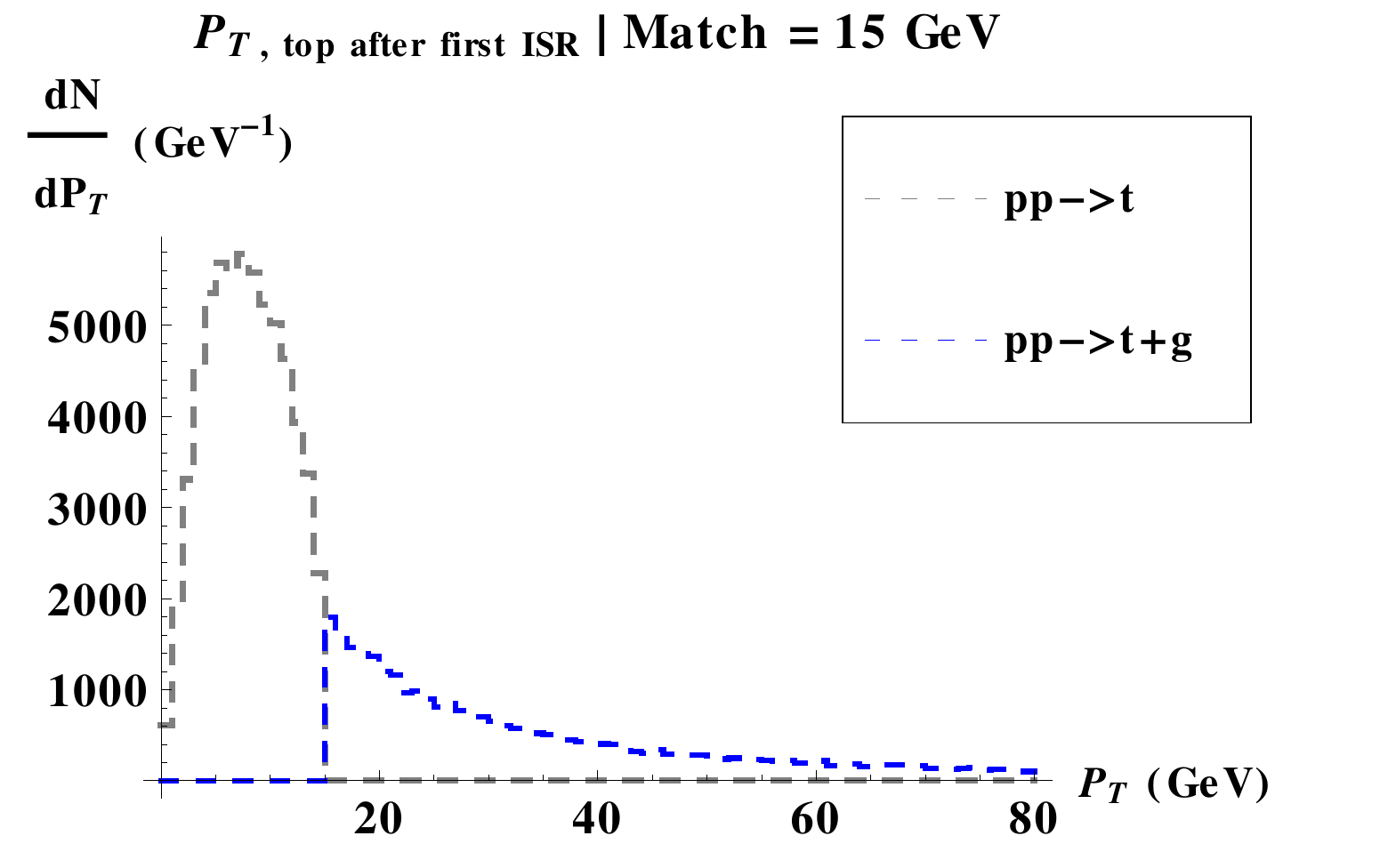}
\caption{$P_T$ distribution of top quark after the first ISR branching with a $P_T^{match} $ of 10 GeV (left)  and 15 GeV (right). }
\label{fig:ptmatch510}
\end{figure}

In figure~\ref{fig:NoMatch} we present the top-quark $P_T$ distribution for direct top 
 after the first emission and with the starting scale shower at $m_t$ (black solid line). The blue dashed line represents the top $P_T$ 
 distribution of the real radiation process. As explained before, both regions overlap and therefore a matching must be 
 introduced in order to avoid double counting. The matching process must be chosen so that the transition between the rescaled 
 direct top distribution and the real radiation process is smooth. In figure~\ref{fig:ptmatch510} we show the
 top $P_T$ distribution for $P_T^{match} = 10,15$ Gev where direct top is the gray dashed line and $t+g$ is the blue dashed line. 
The first one corresponds to the top distribution after the first ISR emission, where the PT-shower starting scale was set to 10 and 15 Gev, respectively; 
the second one is the top distribution from the 't+g' process where a  $P_T^{cut}$ of 10 and 15 Gev was imposed. 
We have studied several $P_T^{match}$ scenarios in the 5 Gev to 20 Gev range  finding no significant differences. We therefore recommend that this matching 
value should be included in any analysis as a systematic error. 

\begin{figure}[h!]
\centering
\hspace{-1.cm}\includegraphics[width=3.0in,angle=0]{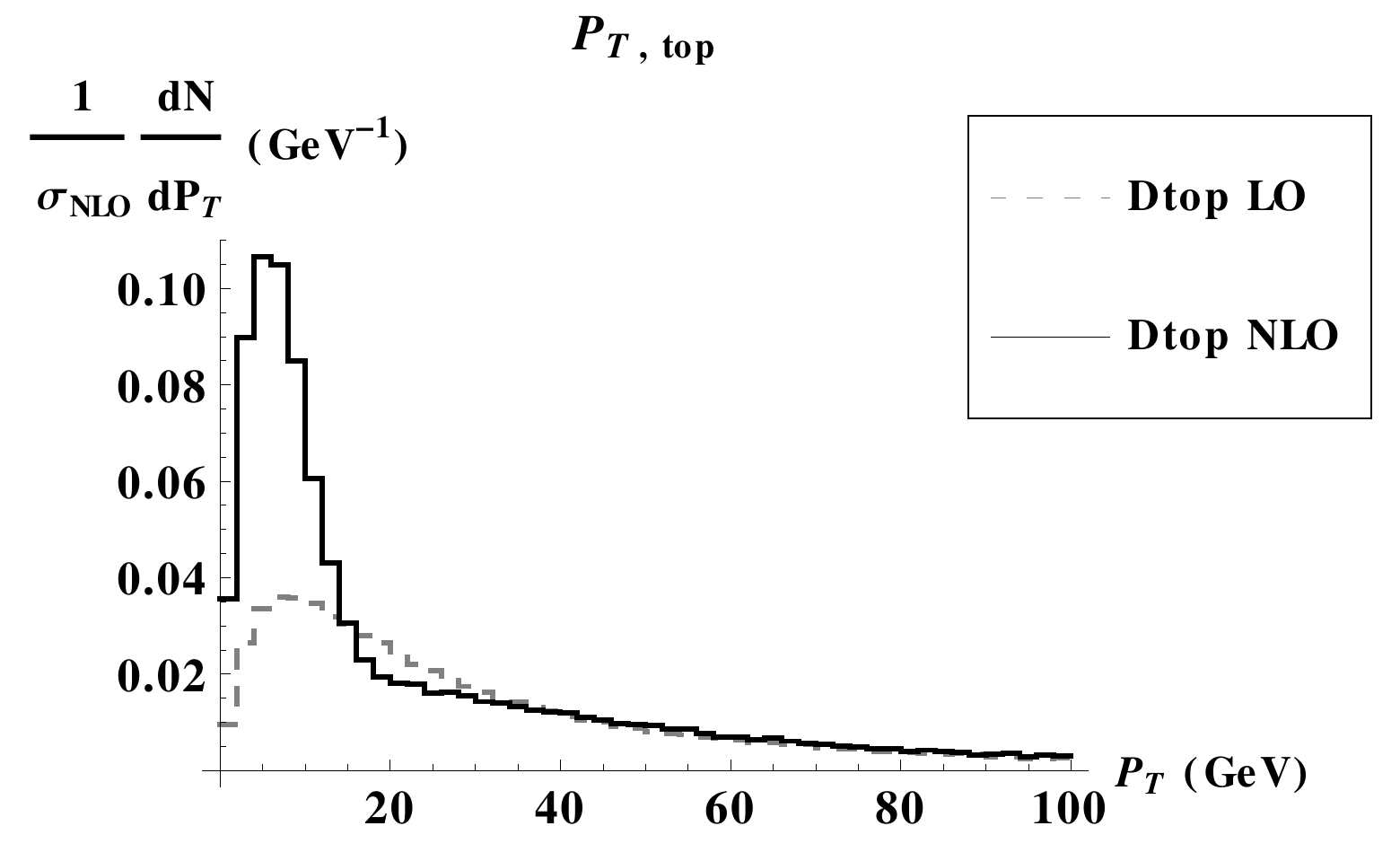}
\hspace{-.3cm}
\includegraphics[width=3.0in,angle=0]{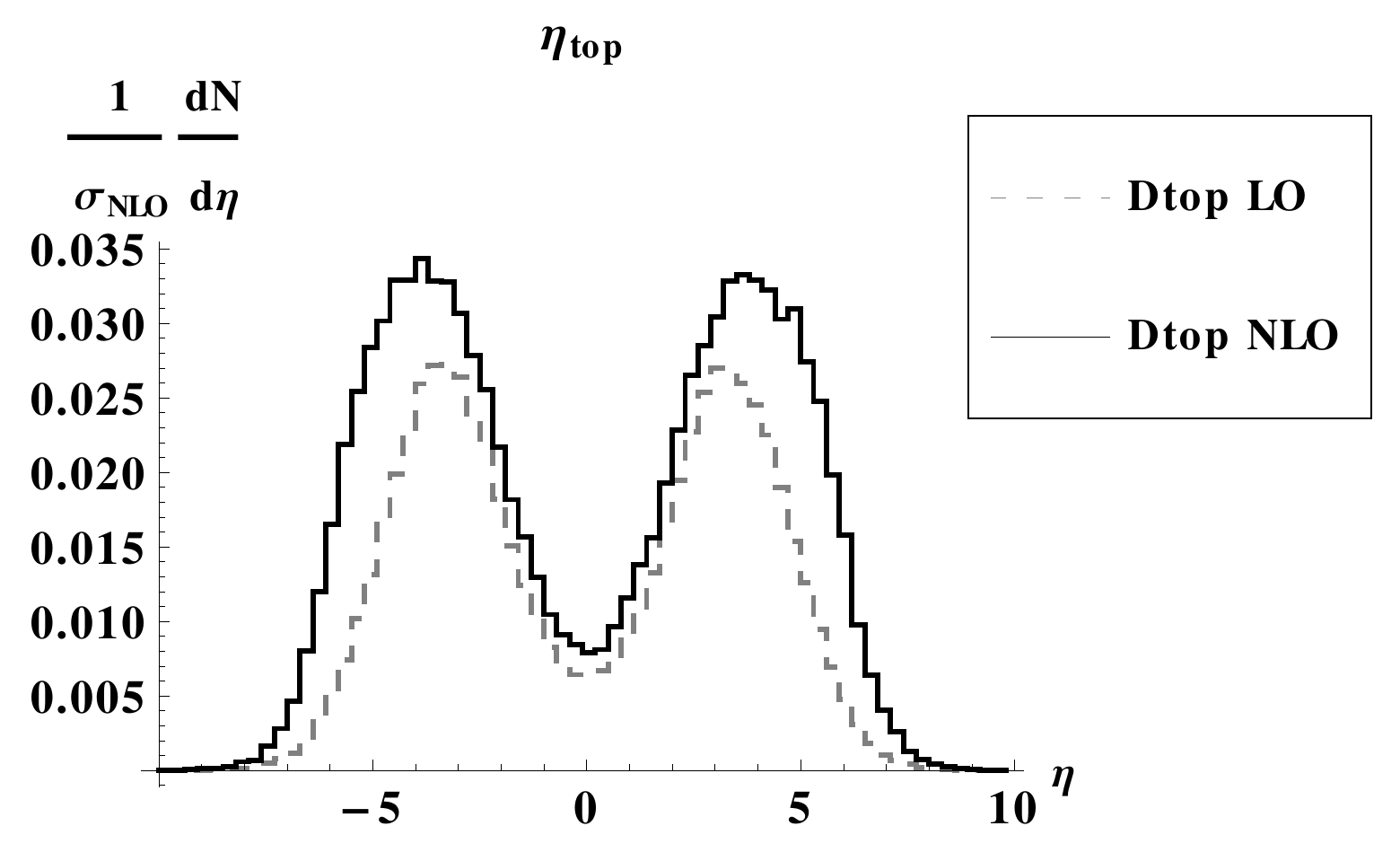}
\caption{Comparison of the LO and NLO $P_T$ (left)  and $\eta$ (right) distributions of the top quark at the partonic level after the full shower (ISR+FSR) and Multiple Interaction. }
\label{fig:LOvsNLOtop}
\end{figure}

\begin{figure}[h!]
\centering
\hspace{-1.cm}\includegraphics[width=3.0in,angle=0]{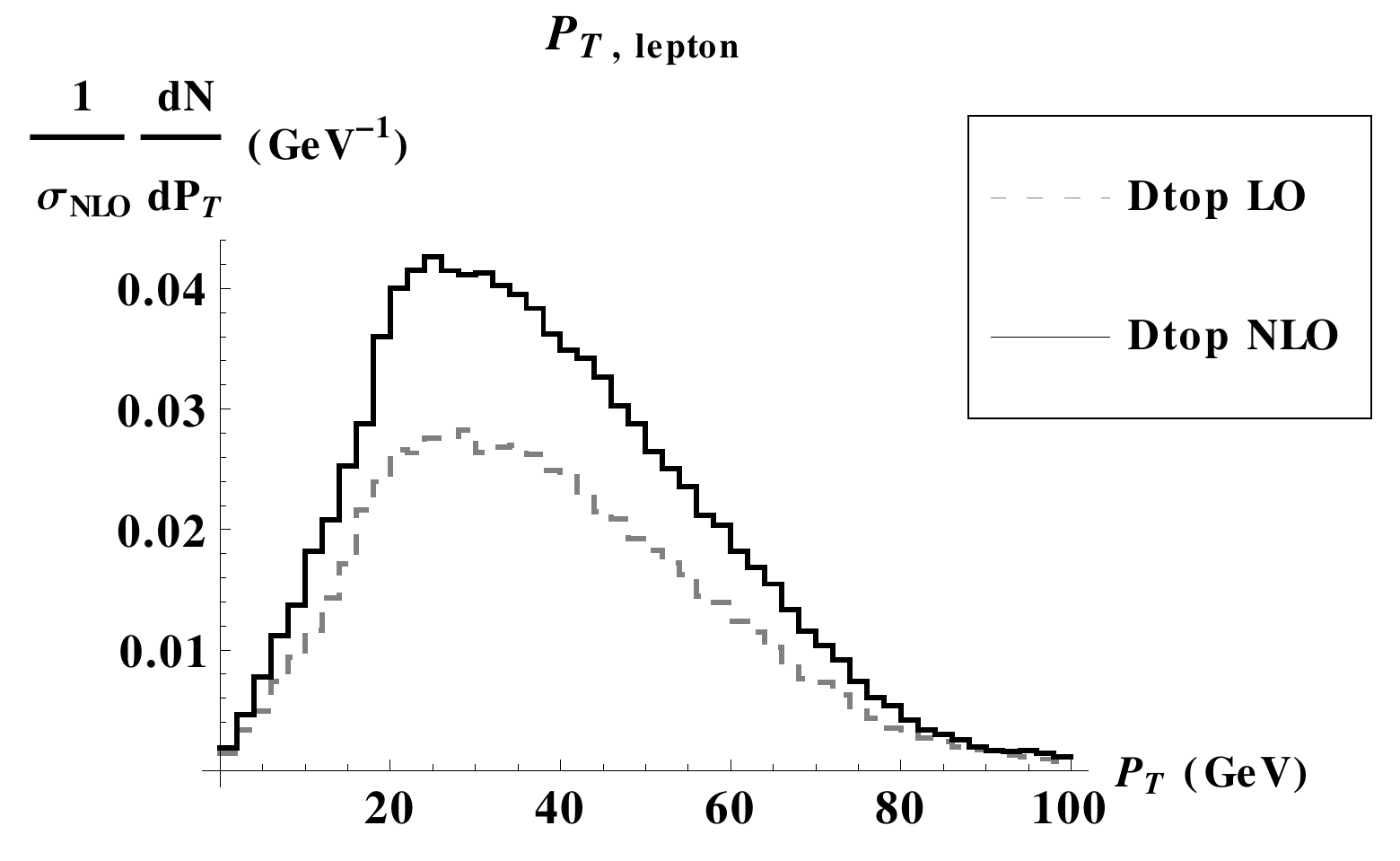}
\hspace{-.3cm}
\includegraphics[width=3.0in,angle=0]{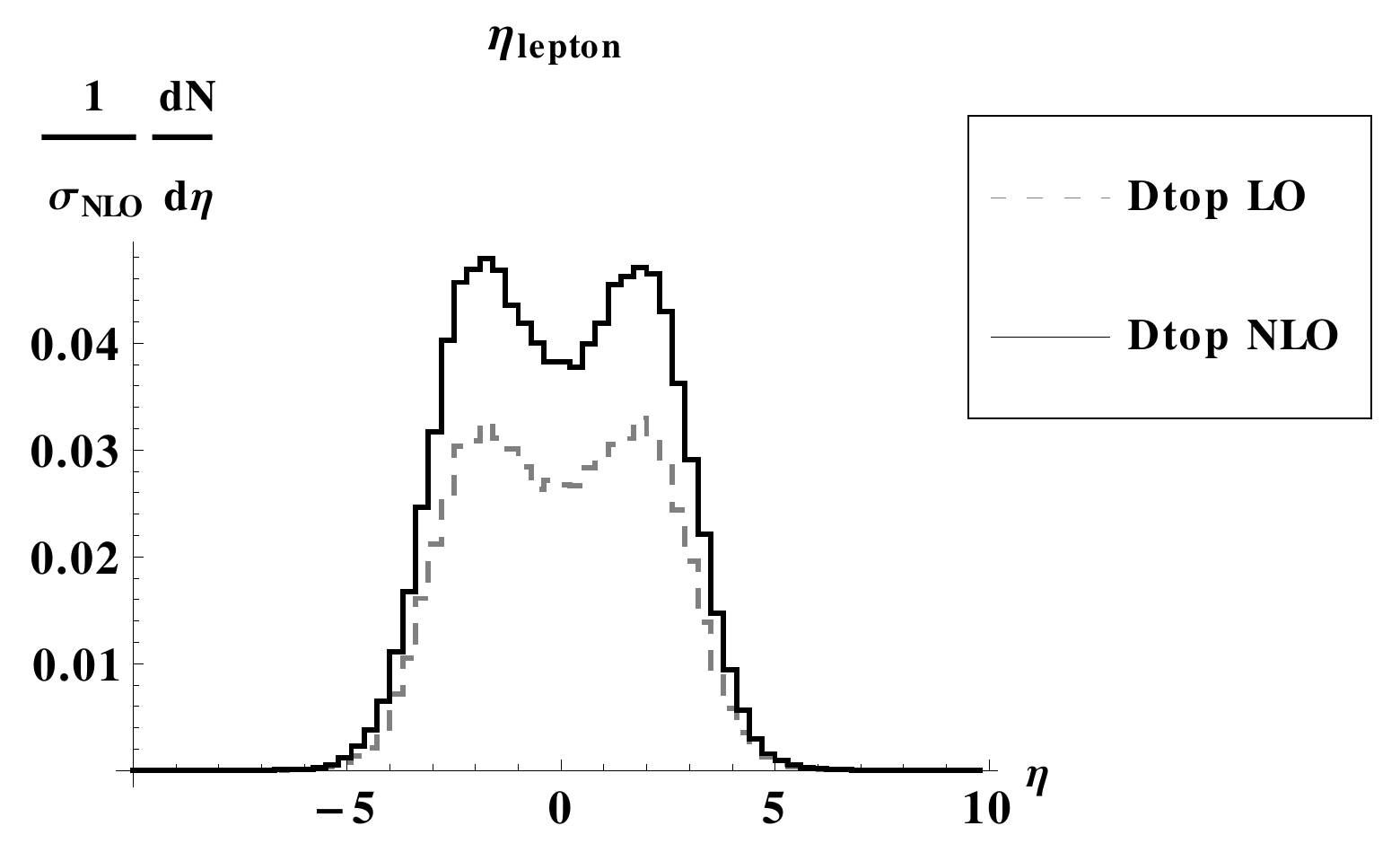}
\caption{Comparison of the LO and NLO $P_T$ (left)  and $\eta$ (right) distributions of the lepton from $t \to bW \to b l \nu$ at the partonic level after the full shower (ISR+FSR) and Multiple Interaction. }
\label{fig:LOvsNLOlepton}
\end{figure}

\begin{figure}[h!]
\centering
\hspace{-1.cm}\includegraphics[width=3.0in,angle=0]{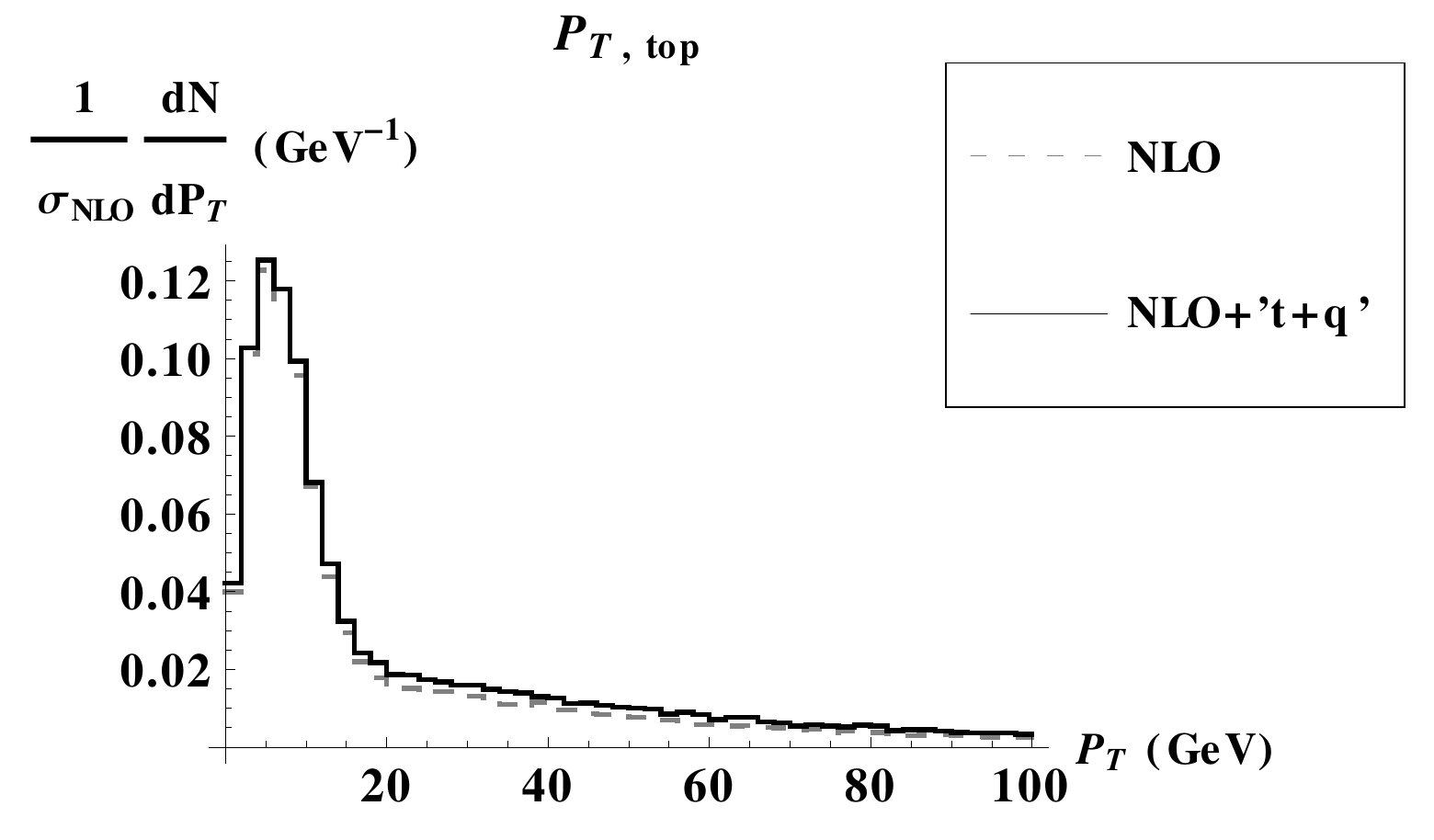}
\hspace{-.3cm}
\includegraphics[width=3.0in,angle=0]{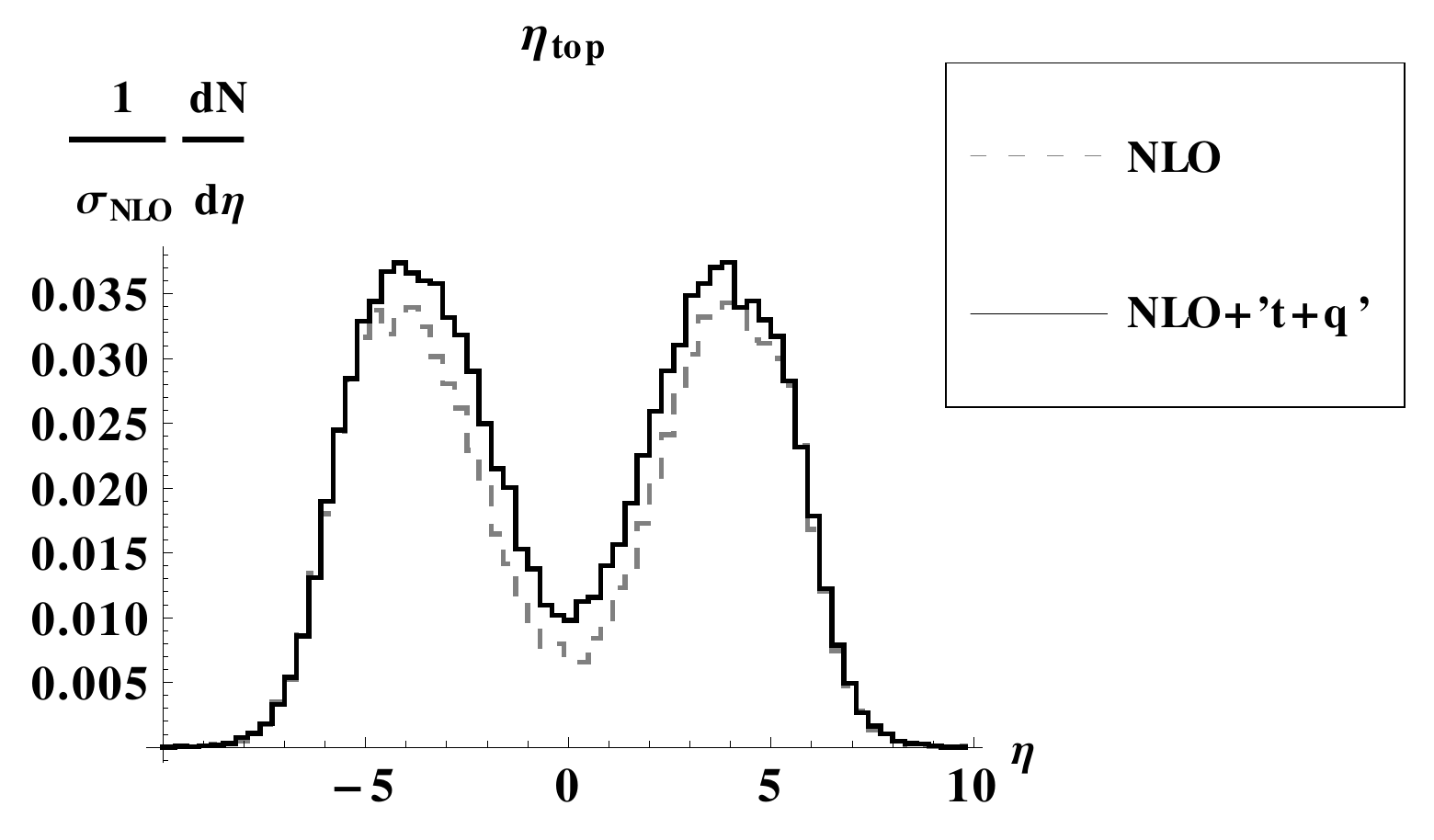}
\caption{$P_T$ (left)  and $\eta$ (right) distributions of the top quark for NLO direct top (solid line)
and NLO direct top plus $pp \to t q$ with $P_T^{match} = 10$ GeV and  jet $p_T > 10$ GeV. }
\label{fig:ptdtopnloq}
\end{figure}

In figure \ref{fig:LOvsNLOtop} we show the final NLO direct top quark $P_T$ and $\eta$ distribution (solid line), as well as the LO result (dashed line)
after full ISR, FSR and Multiple Interactions (MP). It is clear that in both distributions the final NLO result does not amount to simply multiplying the LO
result by a K-factor. In fact, the distributions show that the NLO contribution concentrates more events in the low $P_T$ region and produces top quarks
 at higher angles. In figure~\ref{fig:LOvsNLOlepton} we show the equivalent distribution for the lepton coming from the decaying W. The $b$-quark and 
 neutrino distributions show similar differences between the NLO and LO results. For the top decay products,  the NLO contribution does not result in
 a major change in distributions making a K-factor adjustment suitable. 
  However, one must pay special attention to analysis where the top quark momentum is reconstructed. For example, analysis which include the study of the top 
  spin polarisation usually requires full reconstruction of its momentum. In this case, figure \ref{fig:LOvsNLOtop} show that NLO events should be considered. 
Finally,  when studying the inclusive NLO direct top production one must add the $t+jet$ (single top FCNC) process as well. The distribution of the sum of the two processes is shown
 in figure~\ref{fig:ptdtopnloq} for the top $P_T$ (right) and the top $\eta$ (left). The FCNC Single top process was generated with a $P_T^{cut} = 10$ Gev 
 and only subprocesses where a FCNC interaction takes place were included. Processes like $u\bar{d} \to t\bar{b}$  which are pure SM single top production were discarded. 
 In tables~\ref{tbl:DtopNLO} and \ref{tbl:Lqtop} we show the total contributions from each LO and NLO processes. They were calculated for $k_{u,c}=0.01$ and for LHC @ 7 TeV. 
As we can see from figure~\ref{fig:ptdtopnloq}, the FCNC single top contribution does not produce   considerable shape modifications, although its contribution to the total 
cross section is approximately 13\%.

\begin{table}[h]
\caption{\label{tbl:DtopNLO}FCNC Direct top (anti-top) LO and NLO cross sections for $k_{u,c}=0.01$ and LHC @ 7 Tev.}
\begin{center}
\lineup
\begin{tabular}{lll}
\br
  Subprocess & LO & NLO\\
\mr  
 $ug \to t$ & 6.12 & 8.74  \\
 $cg \to t$ & 0.91 &  1.67  \\ 
 \br
 \vspace{-.3cm}
\end{tabular}
\end{center}
\end{table}

\begin{table}[h]
\caption{\label{tbl:Lqtop}FCNC $t+q$ with $k_{u,c}=0.01$ and LHC @ 7 Tev and $PT_{cut}=10$ Gev.}
\begin{center}
\lineup
\begin{tabular}{lll}
\br
 Subprocess & $k_u = 0.01$ & $k_c = 0.01$\\
\mr  
  $p,p \to t,q$ & 1.12 & 0.40 \\
 \br
  \vspace{-.6cm}
\end{tabular}
\end{center}
\end{table}
 \vspace{-.6cm}

\section{Conclusions}
\label{sec:conclusion}
We have presented a new Monte Carlo generator dedicated to top FCNC physics. We have included the main FCNC top production channels: direct top and single top. 
The direct top channel is available at NLO  and the single top at LO. A complete set of dimension six operators are already included. We have shown that  the NLO top distributions 
obtained for the direct top cannot be built via a K-factor from the LO ones. Therefore NLO events should be used especially in the cases where top-quark momentum 
reconstruction is required. 

\ack{ 
This work is partially supported by the Portuguese
\textit{Funda\c{c}\~{a}o para a Ci\^{e}ncia e a Tecnologia} (FCT)
under contracts CERN/FP/123619/2011
and PTDC/FIS/117951/2010. RS is also partially supported by an 
FP7 Reintegration Grant, number PERG08-GA-2010-277025 and by PEst-OE/FIS/UI0618/2011. 
RC is funded by FCT through the grant SFRH/BPD/45198/2008.}

\end{document}